\documentclass[twocolumn, floatfix,showpacs,preprintnumbers,amsmath,amssymb,pra,superscriptaddress,longbibliography]{revtex4-2}
\usepackage{color}
\usepackage[usenames,dvipsnames,svgnames,table]{xcolor}
\usepackage[colorlinks=true,linkcolor=blue,urlcolor=blue,citecolor=blue]{hyperref}
\usepackage{mathtools}
\usepackage{graphicx}
\usepackage{dcolumn}
\usepackage{array}
\usepackage{lipsum}
\usepackage{bm}
\usepackage{subfigure}
\usepackage{amssymb}
\usepackage{multirow}
\usepackage{tabularx}
\usepackage{amsmath}
\usepackage{braket}
\usepackage{csquotes}
\usepackage{enumitem}
\graphicspath{{plots/}}
 \usepackage{lipsum}
\usepackage{mathrsfs}
\usepackage{MnSymbol}

\usepackage{algorithm}
\usepackage{algpseudocode}
\usepackage{todonotes}

\usepackage[normalem]{ulem} 
\newcommand{\xv}{x}

\begin{document}
\title{The noiseless limit and improved-prior limit of the maximum entropy method and their implications for the analytic continuation problem}

\author{Thomas Chuna}
\email{t.chuna@hzdr.de}
\affiliation{Center for Advanced Systems Understanding (CASUS), D-02826 G\"orlitz, Germany}
\affiliation{Helmholtz-Zentrum Dresden-Rossendorf (HZDR), D-01328 Dresden, Germany}

\author{Nicholas Barnfield}
\affiliation{Department of Statistics, Harvard University, Cambridge, Massachusetts 02138, USA}

\author{Paul Hamann}
\affiliation{Helmholtz-Zentrum Dresden-Rossendorf (HZDR), D-01328 Dresden, Germany}
\affiliation{Institut f\"ur Physik, Universit\"at Rostock, D-18057 Rostock, Germany}

\author{Sebastian Schwalbe}
\affiliation{Center for Advanced Systems Understanding (CASUS), D-02826 G\"orlitz, Germany}
\affiliation{Helmholtz-Zentrum Dresden-Rossendorf (HZDR), D-01328 Dresden, Germany}

\author{Michael P. Friedlander}
\affiliation{Department of Computer Science and Mathematics, University of British Columbia, Vancouver, BC V6T 1Z4, Canada}

\author{Tobias Dornheim}
\affiliation{Helmholtz-Zentrum Dresden-Rossendorf (HZDR), D-01328 Dresden, Germany}
\affiliation{Center for Advanced Systems Understanding (CASUS), D-02826 G\"orlitz, Germany}

\begin{abstract}
Quantum Monte Carlo (QMC) methods are uniquely capable of providing exact simulations of quantum many-body systems. Unfortunately, the applications of a QMC simulation are limited because extracting dynamic properties requires solving the analytic continuation (AC) problem. Across the many fields that use QMC methods, there is no universally accepted analytic continuation algorithm for extracting dynamic properties, but many publications compare to the maximum entropy method. We investigate when entropy maximization is an acceptable approach. We show that stochastic sampling algorithms reduce to entropy maximization when the Bayesian prior is near to the true solution. We investigate when is Bryan's controversial optimization algorithm [Bryan, \textit{Eur.~Biophys.~J.}~\textbf{18}, 165–174 (1990)] for entropy maximization (sometimes known as the maximum entropy method) appropriate to use. We show that Bryan's algorithm is appropriate when the noise is near zero or when the Bayesian prior is near to the true solution. We also investigate the mean squared error, finding a better scaling when the Bayesian prior is near the true solution than when the noise is near zero. We point to examples of improved data-driven Bayesian priors that have already leveraged this advantage. We support these results by solving the double Gaussian problem using both Bryan's algorithm and the newly formulated dual approach to entropy maximization [Chuna et al., \textit{J.~Phys.~A: Math.~Theor.}~\textbf{58}, 335203 (2025)].
\end{abstract}

\maketitle


\section{Introduction}

Much of contemporary physics as well as quantum chemistry and material science is concerned with describing dynamic properties of correlated quantum many-body systems. Quantum Monte Carlo (QMC) methods~\cite{Foulkes_RMP_2001, anderson_QMCtextbook_2007, Ceperley_RevModPhys_1995} are uniquely capable of providing approximation-free simulations of quantum many-body systems that interact via the Coulomb potential. Unfortunately, the applications of QMC simulations are limited in practice because extracting dynamic properties from a simulation is notoriously difficult. Simulations usually generate estimates of $N$-body imaginary-time correlation functions (ITCF)~\cite{boninsegni1, Rabani_PNAS_2002, Dornheim_MRE_2023, Dornheim_JCP_ITCF_2021}, not real-time correlations. Dynamic properties, denoted by spectral function $S(\omega)$, are real-time (or real-frequency) quantities and can only be estimated by analytically continuing imaginary-time correlation functions $F(\tau)$ back to real time~\cite{Jarrell_PhysRep_1996}: 
\begin{eqnarray}
    F(\tau) = \int_{-\infty}^\infty \textnormal{d}\omega\ K(\tau, \omega) \ S(\omega) \label{eq:AC}\ ,
\end{eqnarray}
with $K(\tau,\omega)$ being a known integration kernel. Here, we consider $K(\tau,\omega)=e^{-\tau\omega}$. This kernel relates, for example, the density--density ITCF on the LHS with the dynamic structure factor under the integral; the latter is the key observable, e.g., in x-ray scattering experiments with warm dense matter~\cite{siegfried_review,Dornheim_review, Schoerner_PRE_2023, bespalov_arXiv_2025} and neutron scattering experiments with ultracold atoms~\cite{Filinov_PRA_2012,Dornheim_SciRep_2022,Boninsegni1996, Ferre_PRB_2016}. Additionally, $K(\tau,\omega)=e^{-\tau\omega} \Theta(\omega)$ is typically recovered in the $T=0$ limit of various AC kernels~\cite{Asakawa_PPNP_2001, rothkopf_PhysRep_2020} making it of broader interest.

The task at hand is thus the numerical inversion of (\ref{eq:AC}).
Across the many fields that use QMC methods, there is no universally best analytic continuation (AC) algorithm for extracting dynamic properties. Many methods have been developed for the AC problem. Three main camps are: fitting parameterized models~\cite{Yoon_PRB_2018, Fournier_PRL_2020, xie_DCDS-ABNueralNetwork_2021, wang_PRD_2022, wang2023unsupervised, aarts2025physics, groth_PRB_2019, shi_CPC_2023, dornheim_PRL-DLFC_2018} (this includes neural nets, which are models with an enormous number of parameters), stochastic optimization~\cite{Vitali_PRB_2010, prokofev_JETP_2013, Saccani_Supersolid_PRL_2012, Nichols_PRE_2022, bao_PRB_2016, ShaoSandvik_PhysRep_2023, shu_arXiv_2015, Mishchenko_PRB_2000}, and regularized optimization~\cite{Otsuki_PRE_2017, Otsuki_JPSJ_2020, robles_CPC_2025}. The most ubiquitous method is regularized optimization for example, solving the maximum entropy method (MEM)~\cite{Jarrell_PhysRep_1996, Asakawa_PPNP_2001, Boninsegni_LowTempPhys_1996, Kora_PRB_2018}
\begin{align} \label{eq:GLSEntropy}
    \max_x \quad  -\chi^2[x\mid b] + \alpha S_{SJ}[ x \mid \mu].
\end{align}
Here \eqref{eq:AC} has been discretized such that $A_{ij} = K(\tau_i,\omega_j)$, $x_j = S(\omega_j)$ and $b_i = F(\tau_i)$. The chi-squared goodness-of-fit metric $\chi^2[x\mid b]=\frac{1}{2} \Vert A x - b \Vert_{C^{-1}}$ weights the residual between the QMC data for the ITCF and a proposed solution $x$ by the statistical error of the data $C$. However, the condition number of $A$ is large so the goodness-of-fit metric is not sufficiently constraining and the Shannon-Jaynes entropy $S_{SJ}[ x \mid \mu] = \sum_i \xv_i - \mu_i - \xv_i \ln \xv_i/\mu_i$ weighted by the regularization parameter $\alpha$ penalizes solutions that deviate from the prior $\mu$. Together, these terms create a well-posed strictly convex problem where $x$ is constrained to be positive. 


When new methods are developed, the publications typically contain comparisons with solving \eqref{eq:GLSEntropy}; we emphasize parts of this extensive literature. \textbf{Firstly,} solving equation \eqref{eq:GLSEntropy} is often a limiting case of a more general method. Specifically, Shi \textit{et al}.~\cite{shi_CPC_2023} shows that \eqref{eq:GLSEntropy} is the limit of a single layer neural network, Benedix-Robles \textit{et al}.~\cite{robles_CPC_2025} shows that \eqref{eq:GLSEntropy} is the Dirac delta limit of kernel methods, and Beach~\cite{beach_arXiv_2004} shows that \eqref{eq:GLSEntropy} follows from a ``mean-field'' expansion of stochastic methods. \textbf{Secondly,} the use of \eqref{eq:GLSEntropy} is often debated because practitioners traditionally rely on Bryan's modified Levenberg-Marquardt optimization algorithm~\cite{bryan_EuroBiophys_1990, gull_MaxEntBayesianMethods_1989}, which is controversial because it neglects singular value decomposition (SVD) basis vectors corresponding to small singular values~\cite{rothkopf_Data_2020, rothkopf_PhysRep_2020, asakawa_arXiv_2020, shi_CPC_2023}. Rothkopf~\cite{rothkopf_arXiv_2012, kelly_PRD_2018} has explored ways to improve the basis, but recently Chuna \textit{et al}.~\cite{chuna_JPA_2025} have provided an algorithm that can keep all the basis vectors at a reduced computational cost by solving the dual optimization problem. \textbf{Thirdly,} and likely related to the basis issues, Bryan's algorithm typically produces smoother solutions when compared to stochastic algorithms~\cite{beach_arXiv_2004, rothkopf_arXiv_2012, fuchs_PRE_2010, huang2023acflow}. Such comparisons give an intuition of the approach's mean squared error (MSE), but there is no clear methodological victor~\cite{shu_arXiv_2015, Goulko_PRB_2017, bergeron_PRE_2016} and very few MSE formulas are developed in the literature~\cite{gunnarsson_PRB-MEM_2010}. 
\textbf{Fourthly,} publications are considering the noiseless limit as the best approach to analytic continuation~\cite{ShaoSandvik_PhysRep_2023, Otsuki_PRE_2017, Goulko_PRB_2017, Kora_PRB_2018, Asakawa_PPNP_2001, fei_PRL_2021, kim_JHEP_2018, xie_DCDS-ABNueralNetwork_2021}. Generally, these publications find that as noise decreases Bryan's algorithm performs equally well as other numerical methods (\textit{e.g.} neural nets~\cite{Yoon_PRB_2018, xie_DCDS-ABNueralNetwork_2021, Fournier_PRL_2020}, stochastic sampling and averaging~\cite[Figures 6, 7, 46]{ShaoSandvik_PhysRep_2023}, and dual Newton MEM~\cite{chuna_JPA_2025}).


In this work, \textbf{we investigate when solving \eqref{eq:GLSEntropy} constitutes an acceptable alternative to using stochastic methods.} We find that when the Bayesian prior is near the true solution then Beach's mean-field approximation~\cite{beach_arXiv_2004} is fulfilled and \eqref{eq:GLSEntropy} is valid; this is the first time it has been demonstrated that Beach's approximation is satisfied by a ``real-world'' scenario. \textbf{Additionally, we investigate when solving \eqref{eq:GLSEntropy} via Bryan's algorithm is acceptable.} We find that when the noise is near zero or the Bayesian prior is near the true solution then Bryan's algorithm constitutes an appropriate choice. This is because the estimator of \eqref{eq:GLSEntropy} is linear in those limits and, per Rothkopf~\cite{rothkopf_Data_2020}, Bryan's controversial algorithm relies on the assumption that the estimator is linear. \textbf{We also investigate whether practitioners get a better return on investment improving the Bayesian prior or reducing the noise.} We develop MSE formulas of \eqref{eq:GLSEntropy} in these limits and find that the MSE in the noiseless limit scales with a numerically infinite coefficient, while the MSE for the improved-prior limit does not. We take this to imply that practitioners get a better return-on-investment from improving the Bayesian prior. \textbf{We support the analytic arguments listed above with a numerical investigation of the double Gaussian problem suggested by Goulko \textit{et al}.}~\cite{Goulko_PRB_2017}, applying the dual Newton algorithm and Bryan's algorithm to the problem. 

This paper is organized as follows. Section~\ref{sec:MSE} formulates the mean squared error of \eqref{eq:GLSEntropy} in the noiseless and improved-prior limits and numerically investigates synthetic data generated from the double peak Gaussian problem~\cite{Goulko_PRB_2017}. Section~\ref{sec:MFTexpansion} shows that the improved-prior limit implies the mean-field assumption that Beach used to reduce stochastic methods to entropy maximization. Section~\ref{sec:conclusions} presents the conclusions.

\section{The mean squared error of the maximum entropy estimate}\label{sec:MSE}

\subsection{Definition of the mean squared error in terms of the bias and variance}
The mean squared error (MSE) expresses, on average, the 2-norm distance of your estimator from the true solution. The MSE can be expressed as a bias-variance decomposition and, for a one-dimensional estimator, derivations are commonplace in introductory textbooks~\cite{james2013statisticallearning}. A multi-variate estimator $\hat{\xv}$ has a bias-variance decomposition given by summing over its dimensions. This is expressed as
\begin{subequations}\label{eq:MSE}
\begin{align}\label{eq:biasvariancedecomposition}
    \text{MSE} \equiv \mathbb{E} \left[ \Vert \xv_0  -\hat{\xv} \Vert^2_2 \right] &= \left\Vert \text{Bias}(\hat{\xv}) \right\Vert_2^2 + \text{Tr} \left\{ \text{Cov}(\hat{\xv}) \right\}.
\end{align}
The first term is essentially sum of squares over the component-wise bias,
\begin{align}\label{eq:bias}
    \text{Bias}(\xv_i) = \mathbb{E} \left[ (\xv_0)_i - \hat{\xv}_i \right],
\end{align}
where $x_0$ is the true solution. The second term is the variance, given by the trace of the covariance matrix,
\begin{align}\label{eq:cov}
    \text{Cov}(\hat{\xv})_{ij} = \mathbb{E}\left[ (\hat{\xv}_i - \mathbb{E}\left[ \hat{\xv}_i  \right] ) ( \hat{\xv}_j - \mathbb{E}\left[ \hat{\xv}_j \right] ) \right].
\end{align}
\end{subequations}
The bias-variance decomposition breaks down the MSE into two components: the systematic error and the statistical error. Some additional commentary may be found online~\cite{yeh2019biasvariance, meyer2020biasvariance}.

To use expressions \eqref{eq:bias} and \eqref{eq:cov}, we must have a closed form of the estimator $\hat{x}$. However, the optimum of the entropic regularized cost function~\eqref{eq:GLSEntropy} is given by~\cite{beach_arXiv_2004, gunnarsson_PRB-MEM_2010} 
\begin{align}\label{eq:estimator}
    A^\top C^{-1} A \, \hat{\xv} - A^\top C^{-1} b  +  \alpha \ln \hat{\xv} /\mu = 0
\end{align}
and, as pointed out by Rothkopf~\cite{rothkopf_Data_2020}, the term $\ln \hat{\xv} /\mu$ prevents us from isolating the $\hat{\xv}$ and linearly parameterizing the solution; expansions are needed. Previous investigations of the MSE expanded the solution  $\hat{\xv} = x_0 + \Delta \hat{\xv}$ about the true solution $\xv_0$~\cite{gunnarsson_PRB-MEM_2010}; we do not take that approach here. Instead we consider, for fixed regularization weight $\alpha$, the noiseless limit of uncorrelated data with equal variance, \textit{i.e.} $C^{-1} \approx \sigma^{-2} I$, where
\begin{align}\label{eq:noiseless-approx}
    \sigma^2 \ll 1 
\end{align}
and the limit that the Bayesian prior is near the true solution, which we refer to as the ``improved-prior'' limit
\begin{align}\label{eq:improvedprior-approx}
    \frac{\vert \hat{\xv}_i - \mu_i \vert}{\mu_i} \ll 1.
\end{align}

\subsection{MSE in the noiseless limit}
To compute the MSE, we insert \eqref{eq:noiseless-approx} into \eqref{eq:estimator} yielding
\begin{align}
    \frac{1}{\sigma^2} (A^\top A \, \hat{x} - A^\top b ) +  \alpha \ln \hat{x} /\mu = 0.
\end{align}
In the noiseless limit, the first term dominates. We drop the logarithmic term to recover the ordinary least squares (OLS) estimate. By the Gauss-Markov theorem, this is the best linear un-biased estimator (BLUE), with variance
\begin{align}
\, \text{Cov} (\hat{\boldsymbol{x}}) &= \sigma^2   (A^\top A)^{-1} .
\end{align}
We use the singular value decomposition, $A = U \Sigma V^\top$ and cyclicity of the trace to arrive at
\begin{align}\label{eq:cov-noiseless}
    \text{Tr} \, \text{Cov} (\hat{\boldsymbol{x}}) = \sigma^2 \, \text{Tr} \left\{ \Sigma^{-2} \right\}, 
\end{align}
which exposes how the singular values affect the MSE.

\subsection{MSE in the improved-prior limit}
Evaluating \eqref{eq:MSE} for an improved-prior \eqref{eq:improvedprior-approx} requires a more detailed treatment. To start, we must compute $\hat{\xv}$ and $\mathbb{E}\left[\hat{\xv} \right]$. Beginning from~\eqref{eq:estimator}, we rewrite $\hat{\xv}/\mu = 1 + (\hat{\xv} - \mu)/ \mu$ and linearize the logarithm by Taylor expanding
\begin{align}\label{eq:linearized-SJentropy}
    A^\top C^{-1} A \, \hat{\xv} - A^\top C^{-1} b  +  \alpha \frac{\hat{\xv} - \mu}{\mu} \approx 0.
\end{align}
This estimator resembles the estimator arising from the common $L_2$-distance regularization, but with $x-\mu$ weighted by $1/\mu$. Next we substitute $b = b_0 + \varepsilon = A \xv_0 + \varepsilon$ and insert $0 = A^\top C^{-1} A (\mu - \mu)$. After simplification, this produces 
\begin{subequations}  \label{eq:improved-prior-estimator}
\begin{align}
    \hat{\xv}  =  \mu + M \, A ( \xv_0 -\mu) + M  \varepsilon,
\end{align}
where 
\begin{align}
    M=\left( A^\top C^{-1} A  + \alpha \, \text{diag}(\frac{1}{\mu} ) \right)^{-1} A^\top C^{-1}.
\end{align}  
\end{subequations}
Assuming $\mathbb{E}\left[\varepsilon \right] = 0$ and taking the expectation value yields
\begin{align} \label{eq:improved-prior-expectation-val}
    \mathbb{E}\left[\hat{x} \right] = \mu + M \, A ( x_0 -\mu).
\end{align}


We may now compute the bias \eqref{eq:bias} by substituting equations \eqref{eq:improved-prior-estimator} and \eqref{eq:improved-prior-expectation-val}. We arrive at the expression
\begin{align}
    \text{Bias}\left[ \hat{x} \right] &= \left\Vert \left(I - M \, A \right) ( x_0 -\mu) \right\Vert^2_2. \label{eq:bias-perfectprior}
\end{align}
Grouping $H = M \, A$ and distributing terms yields 
\begin{align}\label{eq:bias-improvedprior}
    \text{Bias}\left[ \hat{x} \right]  = \left\Vert  x_0 -\mu \right\Vert^2_2 -2 \left\Vert  x_0 -\mu \right\Vert^2_H + \left\Vert  x_0 -\mu \right\Vert^2_{H^\top H}.
\end{align}
To expose how the singular values affect the bias, we assume $C = \sigma^2 I$ and simplify via the SVD $A = U \Sigma V^\top$:
\begin{align}
    H =& \, \sigma^{-2} V \left( \sigma^{-2} \Sigma^2 + \alpha \, V^\top \text{diag}(\frac{1}{\mu} ) V \right)^{-1} \Sigma^2 V^\top,
    \\ H^\top H =& \, \sigma^{-2} V  \Sigma^2 \left( \sigma^{-2} \Sigma^2 + \alpha \, V^\top \text{diag}(\frac{1}{\mu} ) V \right)^{-2} \Sigma^2 V^\top.
\end{align}
Notice that for $\sigma \ll 1$ we can drop the $\alpha$ terms, yielding $H = H^\top H = I$ and $\text{Bias}\left[ \hat{x} \right]=0$ as is expected.

Next we compute the trace of the covariance. First, we substitute \eqref{eq:improved-prior-estimator} into \eqref{eq:cov} and cancel terms, which leaves behind only the noise term:
\begin{align}
    \text{Tr} \, \text{Cov}\left( \hat{x} \right) &= \text{Tr} \left\{ \text{Cov}[ M \varepsilon] \right\}\ .
\end{align}
It is interesting to assume $C = \sigma^2 I$ and simplify via the SVD to expose how the singular values affect the statistical error. We pass $M$ through the Cov operator and use cyclity of the trace to produce
\begin{align}\label{eq:cov-improvedprior}
    \text{Tr} \, \text{Cov}\left( \hat{x} \right) &= \sigma^{2} \text{Tr} \left\{ \Sigma^2 \left( \Sigma^2 + \sigma^2 \alpha \, V^\top \text{diag}\left(\frac{1}{\mu} \right) V \right)^{-2} \right\} \ .
\end{align} 
Notice in the limit that $\sigma \ll 1$ we recover \eqref{eq:cov-noiseless}, which is a nice sanity check. 

\subsection{Numeric Investigation of MSE}
\subsubsection{Description of problem}
We present numeric tests of the MSE. In particular, we study synthetic data produced from the double Gaussian problem presented in Goulko \textit{et al}.~\cite{Goulko_PRB_2017}:
\begin{align}
    x_0 = \sum_{i=1}^2 \frac{c_i}{\sigma_i} e^{-\frac{(\omega-z_i)^2}{2\sigma_i^2}},
\end{align}
where the moments are defined, $c_1=0.62; \sigma_1=0.12; z_1=0.74$ and $c_2=0.41; \sigma_2=0.064; z_2=2.93$, the transformation kernel is defined as
\begin{align}
    A = e^{-\tau \omega}, 
\end{align}
and the grid size is $N_\omega =150$, $\omega \in [4.0/N_\omega, 4.0 ]$ and $N_\tau = 30$, $\tau \in [0, 5]$. A plot of $x_0$ is given in Figure~\ref{fig:doubleGaussian}. This Gaussian mixture problem has been of particular interest to many in the analytic continuation community. We generate $N_s=100$ samples of $b$ by adding Gaussian noise with standard deviation $\sigma_0$ that is scaled to the element $b^0$. This is expressed as
\begin{align}\label{eq:noise}
    b^s_i \sim \, \mathcal{N}(b^0_i, (b^0_i \sigma_0)^2),
\end{align} 
where $b_0 = A \, x_0$. From these samples, we estimate $b_i = \text{Avg}_s(b^s_i)$ and $C_{ij} = \text{Var}_s(b_i^s)/N_s \, \delta_{ij}$. So on average the data error is $\sigma_0 / \sqrt{N_s}$. We select the regularization weight via the $\chi^2$-kink algorithm~\cite{Kaufmann_CPC-anacont_2023}.

To investigate the impact of the Bayesian prior, we make the prior a convex combination of the true solution $\xv_0$ and the uniform prior (equivalently flat) $x_\text{flat}$ via,
\begin{align}\label{eq:prior}
    \mu(c) = (1-c) x_0 + c\, x_\text{flat}, \ c \in [0,1].
\end{align}
Essentially, $c$ parameterizes the line through solution space from the flat model ($c=1$) to the true solution ($c=0$). Note that the randomly sampled data has no guarantee that its minimum is located at the true solution. In theory, for very small $c$, the ITCF data may actually make the result worse. As such, we consider $c \in [0.05, 1.0]$; plots of $\mu(c)$ at $c = 1.0, 0.7, 0.5$ are given in Figure~\ref{fig:doubleGaussian}.

To investigate the noiseless limit $\tilde{\alpha} = \sigma^2 \alpha \rightarrow 0$, we logarithmically vary $\frac{\sigma_0}{\sqrt{N_S}} \in [10^{-6}, 10^{-1}]$. For context, in the literature the noiseless limit is typically considered to begin around $10^{-5}$ while realistic data is considered $10^{-3}$. Additionally, $\alpha$ is not a fixed value, but selected by the $\chi^2$-kink algorithm`\cite{Kaufmann_CPC-anacont_2023}. Therefore, it may be the case that as $\sigma^2 \rightarrow 0$ $\chi^2$-kink selects $\alpha \rightarrow \infty$, violating the approximations made when computing the noiseless limit in Section~\ref{sec:MSE}. However, in practice, this does not happen. For example, for $c=0.5$ and $\sigma_0/\sqrt{N_S} = 10^{-6}, 10^{-5}, 10^{-4}, 10^{-3}, 10^{-2}, 10^{-1}$ and the corresponding values of $\sigma^2 \alpha$ are $10^{-9}, 10^{-8}, 10^{-6}, 10^{-3}, 10^{-2}, 10^{0}$, which monotonically also increase with $\sigma_0$. 

\subsubsection{Numeric Results}
\begin{figure}
    \centering
    \includegraphics[width=\linewidth]{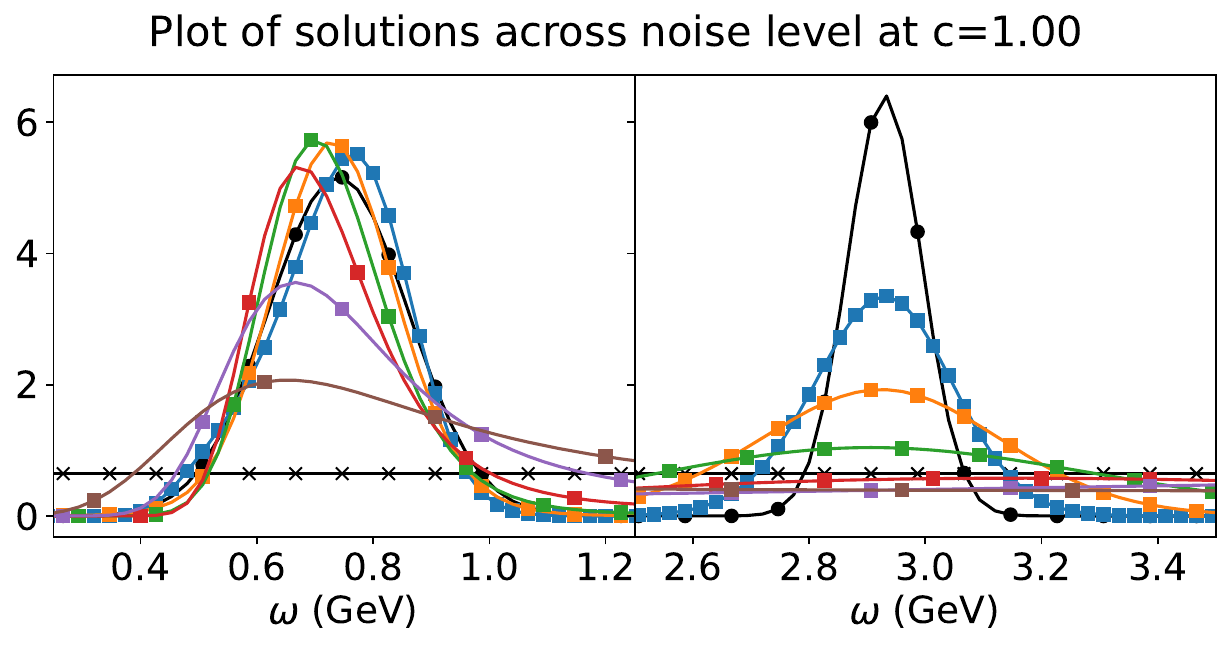}
    \includegraphics[width=\linewidth]{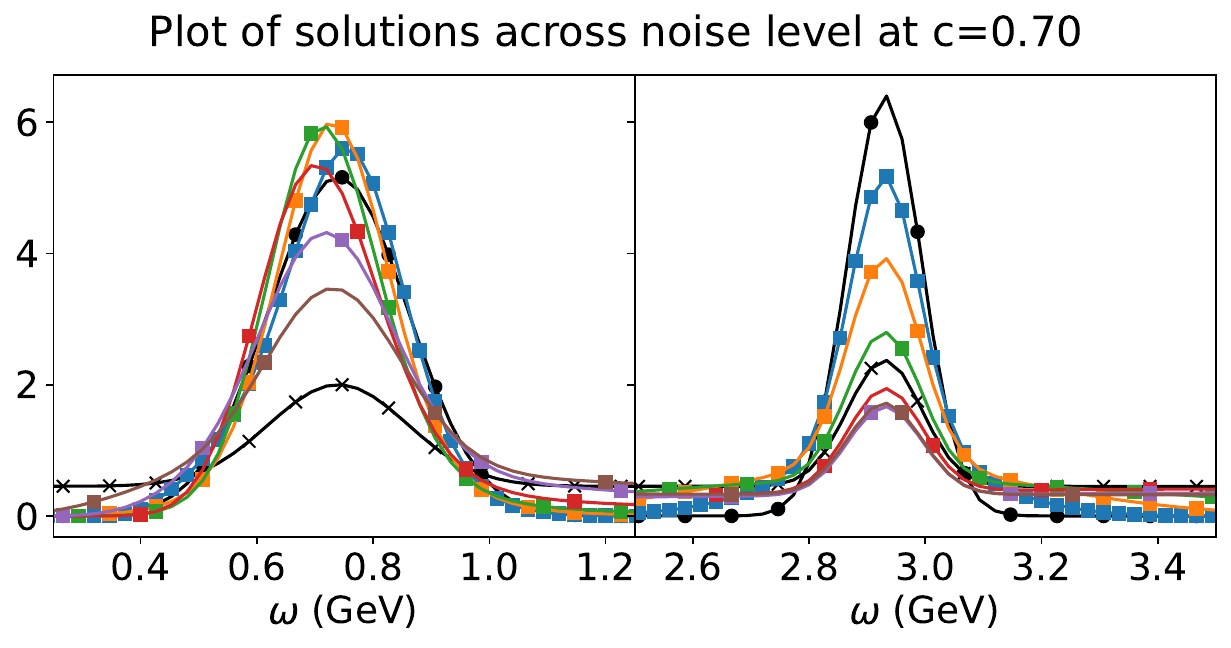}
    \includegraphics[width=\linewidth]{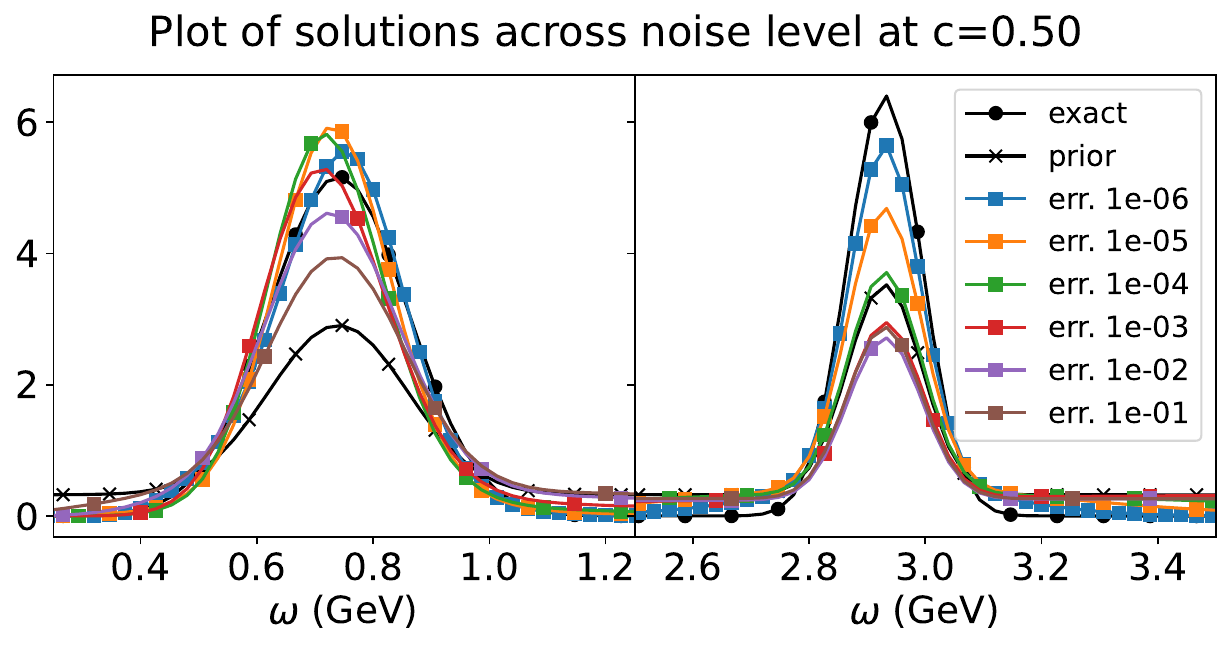}
    \caption{Plot of the double Gaussian test problem from Goulko~\cite{Goulko_PRB_2017} where the left plot is the first peak and the right plot is the second peak. For visualization we neglect the large flat middle region. From top to bottom, we improve the quality of the Bayesian prior, defined \eqref{eq:prior}, with $c=1.0, \, 0.7, \, 0.5 $. Within each plot, we present different noise levels.}
    \label{fig:doubleGaussian}
\end{figure}

We investigate the double Gaussian problem using the dual formulation of the entropic regularization~\cite{chuna_JPA_2025}. Each plot in Figure~\ref{fig:doubleGaussian}, shows the estimate obtained for different quality priors $c=1.0, 0.7, 0.5$. For each prior, we show the estimate converging towards the true solution in the noiseless limit in Figure~\ref{fig:doubleGaussian}. We see that a flat prior $c=1.0$ produces low quality results across all noise levels, but with a small improvement in the prior $c=0.7$ the estimate remains stable at large noise and yields quality results as the noise is reduced; this improvement continues as the prior improves.

\begin{figure}
    \centering
    \includegraphics[width=\linewidth]{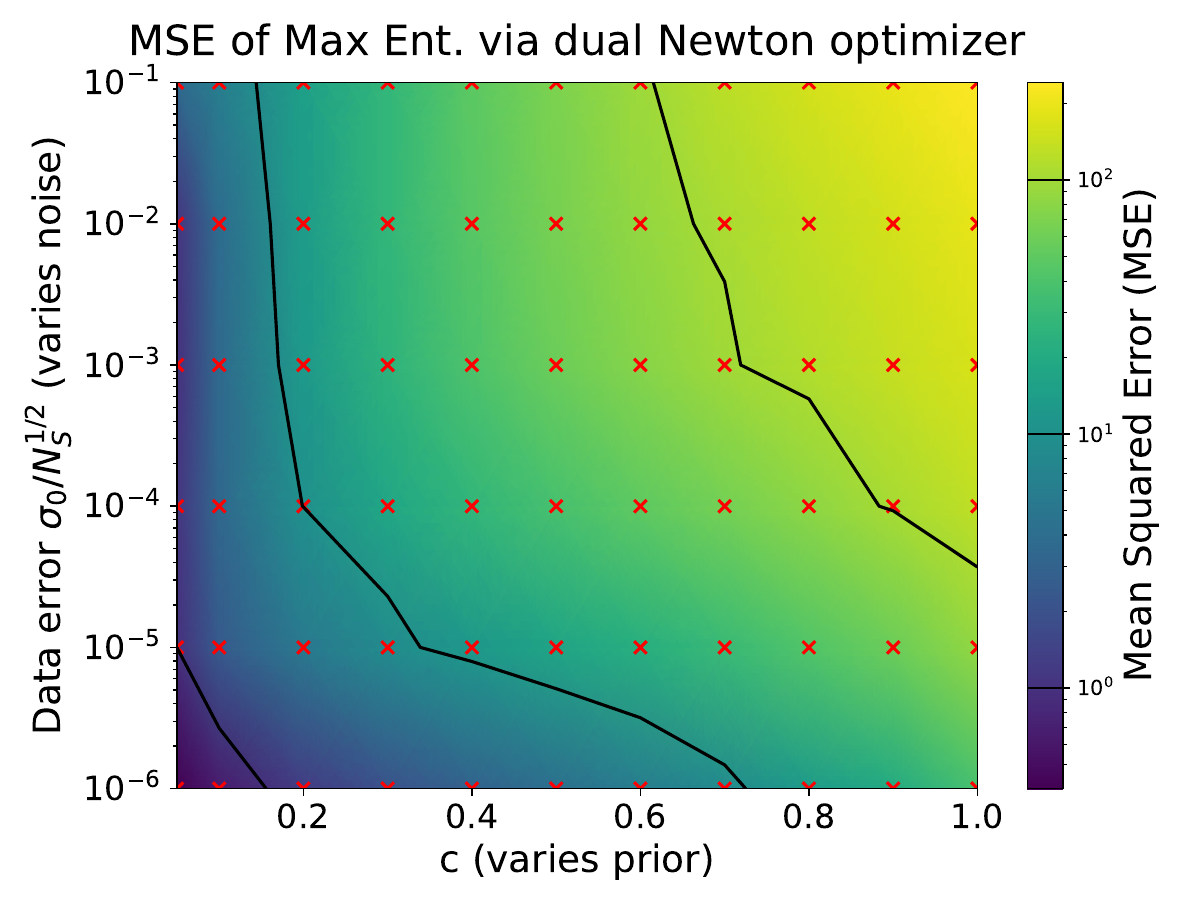}
    \caption{Heatmap of the mean squared error (MSE) over varied noise (noise defined~\eqref{eq:noise}) and prior quality ($c$ defined \eqref{eq:prior}) for the solution produced by the $\chi^2$-kink alg. with a Newton optimizer on the dual problem. Contour levels are also marked as a black horizontal line in the MSE colorbar indicate their values. The grid of red x's indicates the which $c$ and $\sigma_0$ where used, Python's matplotlib smoothly interpolates between these values. \textit{Notice that the y-axis is logarithmic, while the x axis is linear.}}
    \label{fig:MSE}
\end{figure}

Next we investigate the mean squared error $\mathbb{E} \left[ \Vert \xv_0  -\hat{\xv} \Vert^2_2 \right]$. To estimate this quantity, we conduct $N_r=100$ runs of the problem described above for a given $c$ and $\sigma_0$, then average over $N_r$. The MSE estimates are presented in Figure~\ref{fig:MSE}. Since the y-axis is logarithmic and the x axis is linear, improving the default model clearly has a greater impact on the error. This matches the intuition created by the analytic formulas.

\begin{figure}
    \centering
    \includegraphics[width=\linewidth]{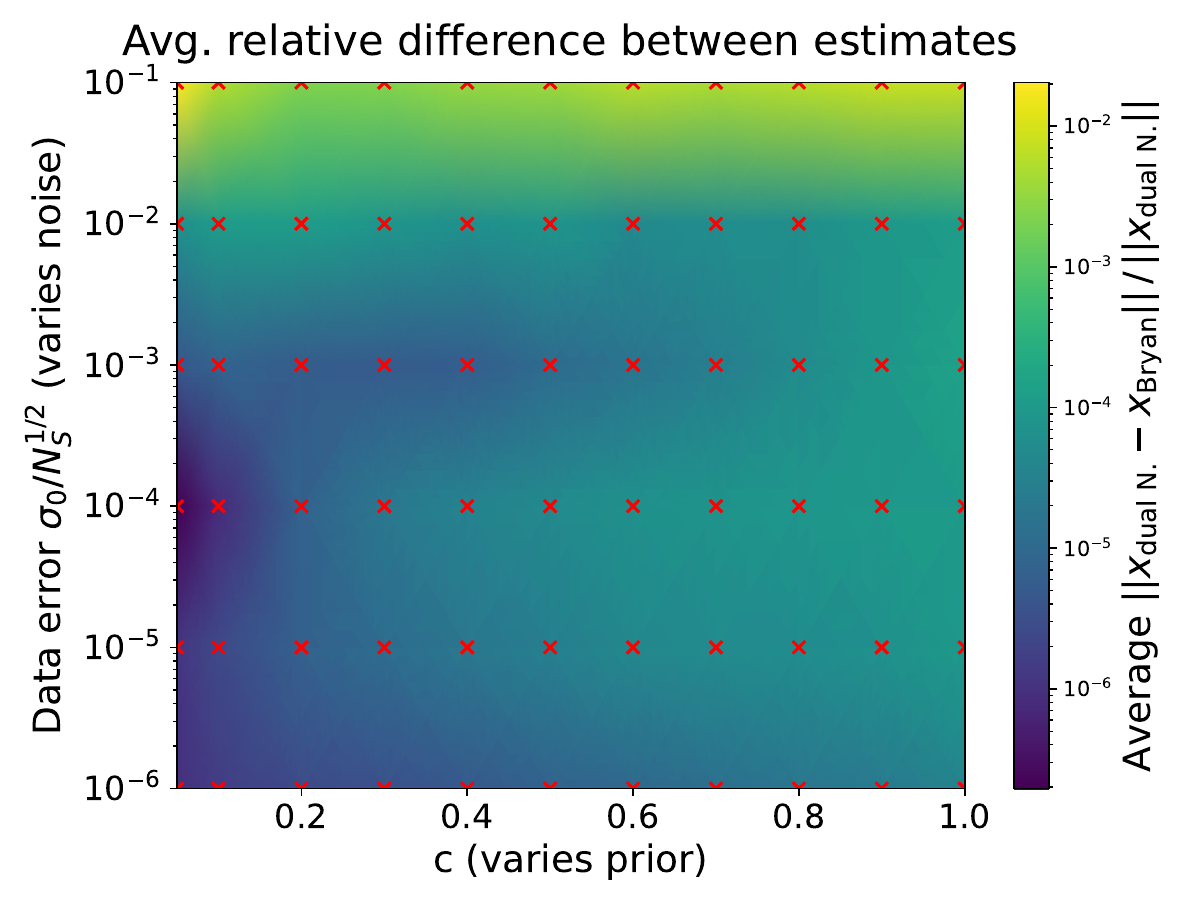}
    \caption{Heatmap of the average relative distance between $\xv_\text{dual N.}$ and $\xv_\text{Bryan}$, which are the solution produced by $\chi^2$-kink alg. with a Newton optimizer on the dual problem and the solution produced by the $\chi^2$-kink alg. with Bryan's modified Levenberg-Marquardt optimizer on the primal problem. In the improved-prior limit (\textit{i.e.} $c \rightarrow 0$) and the noiseless limit the deviations between the estimates go to zero.}
    \label{fig:dual_primal_diff}
\end{figure}

Finally, in Figure~\ref{fig:dual_primal_diff}, we plot a heatmap of the deviation between the solution obtained using Bryan's algorithm and the solution obtained using the dual Newton optimizer. To estimate this deviation, we run both approaches on a given run and compute the 2-norm distance. Then we average over the $N_r=100$ differences to estimate the average deviation. We see that, as expected, in both the noiseless limit and the improved-prior limit there is diminishing deviation between the solution produced using Bryan's algorithm and the solution produced using the dual Newton algorithm.

\section{Identifying MEM as a special case of stochastic methods in the improved-prior limit}\label{sec:MFTexpansion}
Finally, we revisit Beach's seminal work~\cite{beach_arXiv_2004} to show that the improved-prior-limit satisfies Beach's mean-field-approximation.
Beach shows that the MEM cost function is recovered from the mean-field (\textit{i.e.}, saddle point) approximation. In the language of this paper, this amounts to assuming that the covariance~\eqref{eq:cov} scaled by the default model is negligible
\begin{align}\label{eq:BeachMFTexpansion1}
    \mathbb{E} \left[ \left( \frac{\mathbb{E}[\xv_i]}{\mu_i} - \frac{\xv_i}{\mu_i} \right)\left( \frac{\mathbb{E}[\xv_j]}{\mu_j} - \frac{\xv_j}{\mu_j} \right) \right]  \approx 0.
\end{align}
Equation \eqref{eq:BeachMFTexpansion1} is essentially the Ginzburg criterion~\cite[Section 2]{landau_MCstatmech_2021} and the LHS can be rewritten in the typical representation of the covariance as
\begin{align}\label{eq:BeachMFTexpansion2}
    \frac{1}{\mu_i \mu_j} \text{Cov}(\xv_i, \xv_j) \ .
\end{align}
Following Asakawa, Hatsuda, Nakahara's monkey argument~\cite[appendix A]{Asakawa_PPNP_2001}, we assume the covariance between $x_i$ and $x_j$ is described by a multinomial distribution. Essentially, in fixed-grid stochastic sampling there are $M$ blocks to be distributed into $N$ bins. For a particular arrangement of blocks $x$, where $x_i$ is the number of blocks in bin $i$, the probability of a block being in bin $i$ is $p_i = \hat{\xv}_i/M$ ($\mathbb{E}[\xv] = \hat{\xv}$) and the probability that the $i$-th bin receives $x_i$ blocks is given by the multinomial distribution. Thus, the covariance between bins is $\text{Cov}(\xv_i, \xv_j) = - M p_i p_j$, inserting this into \eqref{eq:BeachMFTexpansion2} yields \begin{align}\label{eq:BeachMFTexpansion3}
    \frac{-1}{\mu_i \mu_j}  M^{-1} \hat{x}_i \hat{x}_j  \, .
\end{align}
Notice now that Beach's mean field approximation \eqref{eq:BeachMFTexpansion1} is satisfied by assuming $M \rightarrow \infty$. However, we do not need this assumption in the improved-prior limit, where $\hat{x}/\mu \approx 1$ and \eqref{eq:BeachMFTexpansion3} is reduced to $M^{-1}$. So that even small $M$ values ($\mathcal{O}[10^2]$) satisfy the Ginzberg criterion \eqref{eq:BeachMFTexpansion1}.

Intuitively, we may have expected that the improved-prior limit reduces stochastic sampling to the MEM. In statistical mechanics textbooks, the mean-field approximation is the assumption that the variable of interest has only small perturbations about the mean (for example see Tuckerman \cite[Section 16]{tuckerman2023statistical}). We express this as
\begin{align}\label{eq:physicsMFTexpansion}
    \frac{\xv}{\mu} = \frac{\hat{\xv}}{\mu} + \frac{\delta \xv}{\mu}.
\end{align}
This would precisely be the case if you initialize your stochastic sampling algorithm to a prior $\mu$ near the true solution. Essentially, the samples of $x$ will only be small perturbations about $\mu$ because the sampler will reject large jumps away from this already small $\chi^2$ location.

Finally, we suggest that the noiseless limit does not satisfy Beach's mean-field approximation. Notice in the noiseless limit that, because the least squares estimate is the BLUE then $\hat{x} = x_0$, which does not immediately reduce \eqref{eq:BeachMFTexpansion3} to $M^{-1}$. To add strength to the suggestion, the free term and the interaction term in Beach's seminal paper~\cite{beach_arXiv_2004} (equations 23, 24, and 25) are both scaled by $1/\sigma^2$. Thus, the noiseless limit would not uniquely affect the interaction term and thus would not facilitate the expansion.


\section{Conclusions} \label{sec:conclusions}

Our three major conclusions are: (a) if the Bayesian prior is near the true solution then stochastic sampling methods reduce to solving the entropy regularized least squares problem \eqref{eq:GLSEntropy}; (b) if the Bayesian prior is near the true solution, then the estimator reduces to a linear estimate and thus Bryan's algorithm becomes valid; (c) the best way to reduce the mean squared error of the estimate is to improve the Bayesian prior.

(a), in Section~\ref{sec:MFTexpansion}, we demonstrated that the assumption of an improved-prior~\eqref{eq:improvedprior-approx} is sufficient to reduce stochastic methods to the MEM. Essentially, the improved-prior limit implies Beach's mean-field approximation. This result provides an intuitive situation in which the Beach's assumption is valid. Additionally, we suggest that the noise-less limit does not imply Beach's mean-field approximation.

(b), in Section~\ref{sec:MSE}, we demonstrated that the noiseless and improved-prior limits reduced the MEM estimator $\hat{\xv}$ to a linear estimator. As per Rothkopf~\cite{rothkopf_Data_2020}, this implies that Bryan's null-space trick is valid in these limits. We also observed this to be the case in our numeric investigation, see Figure~\ref{fig:dual_primal_diff}. This result explains a host of results in the literature where MEM performs equally well as other methods in the noiseless limit. The key takeaway being that other algorithms are \textbf{expected} to outperform the MEM when the signal-to-noise ratio is large because Bryan's algorithm is not valid in that limit.

(c), in Section~\ref{sec:MSE}, we have shown that in the noiseless limit, the MSE is $\sigma^2 \, \text{Tr} \, \Sigma^{-2}$. Thus, quadratic improvements may be found with vanishing noise, but for the Laplace kernel $\text{Tr} \, \Sigma^{-2}$ is numerically infinite. Given the extreme cost of quantum Monte Carlo simulations this is not an appealing approach. By comparison, in the improved-prior-limit the bias~\eqref{eq:bias-perfectprior} and the variance~\eqref{eq:cov-improvedprior} contain quadratically vanishing terms that are scaled by matrix $M$, which is windowed by $\alpha \, \text{diag}(1 / \mu)$ (\textit{i.e.} small singular values do not lead to numerical infinity). As a result, you can reap the benefits of the quadratically diminishing terms. The MSE was estimated numerically for the double Gaussian problem to verify these claims, see Figure~\ref{fig:MSE}. 

Future work from the stochastic community may wish to investigate the intuition discussed in Section~\ref{sec:MFTexpansion} near \eqref{eq:physicsMFTexpansion}. Future work from the regularized optimization community may wish to consider the best method for creating an improved Bayesian prior, \textit{e.g.}, preprocess the data to create a prior. Early examples of data driven priors were based on moment matching~\cite{Jarrell_PhysRep_1996, gunnarsson_PRB-MEM_2010, bergeron_PRE_2016}, \textit{i.e.}, using a Gaussian whose moments were computed from the ITCF data. Recent publications~\cite{chuna_PRE_2025, chuna_JCP_2025} developed a data-driven priors that use the static approximation~\cite{dornheim_JCP-MLstatic_2019, dornheim_PRB-analyticstatic_2021} derived from density response theory~\cite{GiulianiVignale_quantumtheory_2008, Dornheim_review}. Within \cite{chuna_PRE_2025} it is argued that using a data-driven prior Bayesian (\textit{i.e.} a simple parameterized model informed by the ITCF data) is similar in spirit to detrending approaches in data science~\cite{wu2007trend, brockwell2002TSA, adhikari2013TSA, james2013statisticallearning}. Additionally, it is detailed how the error arising from data driven priors can be quantified via leave-one-out binning~\cite{berg_book_2004}. We note for the imaginative reader that Gunnarson \textit{et al}.~\cite{gunnarsson_PRB-MEM_2010} attempted to iterate MEM, using its own solution as the Bayesian prior for the next iteration, but this approach failed because systematic error / bias grew too large.


\begin{acknowledgements}

This work was partially supported by the Center for Advanced Systems Understanding (CASUS), financed by Germany’s Federal Ministry of Education and Research (BMBF) and the Saxon state government out of the State budget approved by the Saxon State Parliament.  This work has received funding from the European Union's Just Transition Fund (JTF) within the project \textit{R\"ontgenlaser-Optimierung der Laserfusion} (ROLF), contract number 5086999001, co-financed by the Saxon state government out of the State budget approved by the Saxon State Parliament. This work has received funding from the European Research Council (ERC) under the European Union’s Horizon 2022 research and innovation programme
(Grant agreement No. 101076233, "PREXTREME"). 
Views and opinions expressed are however those of the authors only and do not necessarily reflect those of the European Union or the European Research Council Executive Agency. Neither the European Union nor the granting authority can be held responsible for them. 
Tobias Dornheim gratefully acknowledges funding from the Deutsche Forschungsgemeinschaft (DFG) via project DO 2670/1-1.
Computations were performed on a Bull Cluster at the Center for Information Services and High-Performance Computing (ZIH) at Technische Universit\"at Dresden and at the Norddeutscher Verbund f\"ur Hoch- und H\"ochstleistungsrechnen (HLRN) under grant mvp00024.
\end{acknowledgements}

\bibliography{bibliography}

@book{tuckerman2023statistical,
  title={Statistical mechanics: theory and molecular simulation},
  author={Tuckerman, Mark E},
  year={2023},
  publisher={Oxford university press}
}

@book{landau_MCstatmech_2021,
  title={A guide to Monte Carlo simulations in statistical physics},
  author={Landau, David and Binder, Kurt},
  year={2021},
  publisher={Cambridge university press}
}

@misc{bespalov_arXiv_2025,
      title={Experimental validation of electron correlation models in warm dense matter}, 
      author={Dmitrii S. Bespalov and Ulf Zastrau and Zhandos A. Moldabekov and Thomas Gawne and Tobias Dornheim and Moyassar Meshhal and Alexis Amouretti and Michal Andrzejewski and Karen Appel and Carsten Baehtz and Erik Brambrink and Khachiwan Buakor and Carolina Camarda and David Chin and Gilbert Collins and Celine Crepisson and Adrien Descamps and Jon Eggert and Luke Fletcher and Alessandro Forte and Gianluca Gregori and Marion Harmand and Oliver S. Humphries and Hauke Hoeppner and Jonas Kuhlke and William Lynn and Julian Luetgert and Masruri Masruri and Emma M. McBride and Ryan Stewart McWilliams and Alan Augusto Sanjuan Mora and Jean-Paul Naedler and Paul Neumayer and Charlotte Palmer and Alexander Pelka and Lea Pennacchioni and Danae Polsin and Calum Prestwood and Natalia A. Pukhareva and Chongbing Qu and Divyanshu Ranjan and Ronald Redmer and Michael Roeper and Christoph Sahle and Samuel Schumacher and Jan-Patrick Schwinkendorf and Melanie J. Sieber and Madison Singleton and Ethan Smith and Christian Sternemann and Thomas Stevens and Michael Stevenson and Cornelius Strohm and Minxue Tang and Monika Toncian and Toma Toncian and Thomas Tschentscher and Sam Vinko and Justin Wark and Max Wilke and Dominik Kraus and Thomas R. Preston},
      year={2025},
      eprint={2509.10107},
      archivePrefix={arXiv},
      primaryClass={physics.plasm-ph},
      url={https://arxiv.org/abs/2509.10107}, 
}

@book{GiulianiVignale_quantumtheory_2008,
address = {Cambridge},
author = {G. Giuliani and G. Vignale},
publisher = {Cambridge University Press},
title = {Quantum Theory of the Electron Liquid},
year = {2008},
}

@article{shi_CPC_2023,
  title={Rethinking the ill-posedness of the spectral function reconstruction—Why is it fundamentally hard and how Artificial Neural Networks can help},
  author={Shi, Shuzhe and Wang, Lingxiao and Zhou, Kai},
  journal={Computer Physics Communications},
  volume={282},
  pages={108547},
  year={2023},
  publisher={Elsevier}
}

@article{shu_arXiv_2015,
  title={A stochastic approach to the reconstruction of spectral functions in lattice QCD},
  author={Shu, Hai-Tao and Ding, Heng-Tong and Kaczmarek, Olaf and Mukherjee, Swagato and Ohno, Hiroshi},
  journal={arXiv preprint arXiv:1510.02901},
  year={2015}
}

@article{Mishchenko_PRB_2000,
  title = {Diagrammatic quantum Monte Carlo study of the Fr\"ohlich polaron},
  author = {Mishchenko, A. S. and Prokof'ev, N. V. and Sakamoto, A. and Svistunov, B. V.},
  journal = {Phys. Rev. B},
  volume = {62},
  issue = {10},
  pages = {6317--6336},
  numpages = {0},
  year = {2000},
  month = {Sep},
  publisher = {American Physical Society},
  doi = {10.1103/PhysRevB.62.6317},
  url = {https://link.aps.org/doi/10.1103/PhysRevB.62.6317}
}

@article{beach_arXiv_2004,
  title={Identifying the maximum entropy method as a special limit of stochastic analytic continuation},
  author={Beach, KSD},
  journal={arXiv preprint cond-mat/0403055},
  year={2004}
}

@article{ShaoSandvik_PhysRep_2023,
  title={Progress on stochastic analytic continuation of quantum Monte Carlo data},
  author={Shao, Hui and Sandvik, Anders W},
  journal={Physics Reports},
  volume={1003},
  pages={1--88},
  year={2023},
  publisher={Elsevier}
}

@article{groth_PRB_2019,
  title={Ab initio path integral Monte Carlo approach to the static and dynamic density response of the uniform electron gas},
  author={Groth, Simon and Dornheim, Tobias and Vorberger, Jan},
  journal={Physical Review B},
  volume={99},
  number={23},
  pages={235122},
  year={2019},
  publisher={APS}
}

@article{bao_PRB_2016,
  title={Fast and efficient stochastic optimization for analytic continuation},
  author={Bao, Feng and Tang, Y and Summers, M and Zhang, G and Webster, C and Scarola, V and Maier, Thomas A},
  journal={Physical Review B},
  volume={94},
  number={12},
  pages={125149},
  year={2016},
  publisher={APS}
}

@article{Nichols_PRE_2022,
  title = {Parameter-free differential evolution algorithm for the analytic continuation of imaginary time correlation functions},
  author = {Nichols, Nathan S. and Sokol, Paul and Del Maestro, Adrian},
  journal = {Phys. Rev. E},
  volume = {106},
  issue = {2},
  pages = {025312},
  numpages = {10},
  year = {2022},
  month = {Aug},
  publisher = {American Physical Society},
  doi = {10.1103/PhysRevE.106.025312},
  url = {https://link.aps.org/doi/10.1103/PhysRevE.106.025312}
}

@article{Vitali_PRB_2010,
  title = {Ab initio low-energy dynamics of superfluid and solid $^{4}\textnormal{H}\textnormal{e}$},
  author = {Vitali, E. and Rossi, M. and Reatto, L. and Galli, D. E.},
  journal = {Phys. Rev. B},
  volume = {82},
  issue = {17},
  pages = {174510},
  numpages = {14},
  year = {2010},
  month = {Nov},
  publisher = {American Physical Society},
  doi = {10.1103/PhysRevB.82.174510},
  url = {https://link.aps.org/doi/10.1103/PhysRevB.82.174510}
}

@article{Saccani_Supersolid_PRL_2012,
  title = {Excitation Spectrum of a Supersolid},
  author = {Saccani, S. and Moroni, S. and Boninsegni, M.},
  journal = {Phys. Rev. Lett.},
  volume = {108},
  issue = {17},
  pages = {175301},
  numpages = {5},
  year = {2012},
  month = {Apr},
  publisher = {American Physical Society},
  doi = {10.1103/PhysRevLett.108.175301},
  url = {https://link.aps.org/doi/10.1103/PhysRevLett.108.175301}
}

@article{prokofev_JETP_2013,
  title={Spectral analysis by the method of consistent constraints},
  author={Prokof’ev, Nikolay V and Svistunov, Boris V},
  journal={JETP letters},
  volume={97},
  number={11},
  pages={649--653},
  year={2013},
  publisher={Springer}
}

@article{Yoon_PRB_2018,
  title = {Analytic continuation via domain knowledge free machine learning},
  author = {Yoon, Hongkee and Sim, Jae-Hoon and Han, Myung Joon},
  journal = {Phys. Rev. B},
  volume = {98},
  issue = {24},
  pages = {245101},
  numpages = {7},
  year = {2018},
  month = {Dec},
  publisher = {American Physical Society},
  doi = {10.1103/PhysRevB.98.245101},
  url = {https://link.aps.org/doi/10.1103/PhysRevB.98.245101}
}

@article{Fournier_PRL_2020,
  title = {Artificial Neural Network Approach to the Analytic Continuation Problem},
  author = {Fournier, Romain and Wang, Lei and Yazyev, Oleg V. and Wu, QuanSheng},
  journal = {Phys. Rev. Lett.},
  volume = {124},
  issue = {5},
  pages = {056401},
  numpages = {6},
  year = {2020},
  month = {Feb},
  publisher = {American Physical Society},
  doi = {10.1103/PhysRevLett.124.056401},
  url = {https://link.aps.org/doi/10.1103/PhysRevLett.124.056401}
}

@article{xie_DCDS-ABNueralNetwork_2021,
  title={Analytic continuation of noisy data using Adams Bashforth residual neural network},
  author={Xie, Xuping and Bao, Feng and Maier, Thomas and Webster, Clayton},
  journal={Discrete and Continuous Dynamical Systems-Series S},
  volume={15},
  number={4},
  year={2021},
  publisher={Oak Ridge National Laboratory (ORNL), Oak Ridge, TN (United States)}
}

@article{wang_PRD_2022,
  title={Reconstructing spectral functions via automatic differentiation},
  author={Wang, Lingxiao and Shi, Shuzhe and Zhou, Kai},
  journal={Physical Review D},
  volume={106},
  number={5},
  pages={L051502},
  year={2022},
  publisher={APS}
}

@inproceedings{wang2023unsupervised,
  title={Unsupervised learning spectral functions with neural networks},
  author={Wang, Lingxiao and Shi, Shuzhe and Zhou, Kai},
  booktitle={Journal of Physics: Conference Series},
  volume={2586},
  pages={012158},
  year={2023},
  organization={IOP Publishing}
}

@article{aarts2025physics,
  title={Physics-driven learning for inverse problems in quantum chromodynamics},
  author={Aarts, Gert and Fukushima, Kenji and Hatsuda, Tetsuo and Ipp, Andreas and Shi, Shuzhe and Wang, Lingxiao and Zhou, Kai},
  journal={Nature Reviews Physics},
  volume={7},
  number={3},
  pages={154--163},
  year={2025},
  publisher={Nature Publishing Group UK London}
}

@article{robles_CPC_2025,
  title={PyLIT: Reformulation and implementation of the analytic continuation problem using kernel representation methods},
  author={Robles, Alexander Benedix and Hofmann, Phil-Alexander and Chuna, Thomas and Dornheim, Tobias and Hecht, Michael},
  journal={arXiv preprint arXiv:2505.10211},
  year={2025}
}

@article{Otsuki_PRE_2017,
  title = {Sparse modeling approach to analytical continuation of imaginary-time quantum Monte Carlo data},
  author = {Otsuki, Junya and Ohzeki, Masayuki and Shinaoka, Hiroshi and Yoshimi, Kazuyoshi},
  journal = {Phys. Rev. E},
  volume = {95},
  issue = {6},
  pages = {061302},
  numpages = {6},
  year = {2017},
  month = {Jun},
  publisher = {American Physical Society},
  doi = {10.1103/PhysRevE.95.061302},
  url = {https://link.aps.org/doi/10.1103/PhysRevE.95.061302}
}

@article{Otsuki_JPSJ_2020,
author = {Otsuki ,Junya and Ohzeki ,Masayuki and Shinaoka ,Hiroshi and Yoshimi ,Kazuyoshi},
title = {Sparse Modeling in Quantum Many-Body Problems},
journal = {Journal of the Physical Society of Japan},
volume = {89},
number = {1},
pages = {012001},
year = {2020},
doi = {10.7566/JPSJ.89.012001},
URL = {https://doi.org/10.7566/JPSJ.89.012001},
eprint = {https://doi.org/10.7566/JPSJ.89.012001}
}

@incollection{gull_MaxEntBayesianMethods_1989,
  title={Developments in maximum entropy data analysis},
  author={Gull, Stephen F},
  booktitle={Maximum Entropy and Bayesian Methods},
  pages={53--71},
  year={1989},
  publisher={Springer}
}

@article{bryan_EuroBiophys_1990,
  title={Maximum entropy analysis of oversampled data problems},
  author={Bryan, RK},
  journal={European Biophysics Journal},
  volume={18},
  pages={165--174},
  year={1990},
  publisher={Springer}
}

@article{Jarrell_PhysRep_1996,
title = {Bayesian inference and the analytic continuation of imaginary-time quantum Monte Carlo data},
journal = {Physics Reports},
volume = {269},
number = {3},
pages = {133-195},
year = {1996},
issn = {0370-1573},
doi = {https://doi.org/10.1016/0370-1573(95)00074-7},
url = {https://www.sciencedirect.com/science/article/pii/0370157395000747},
author = {Mark Jarrell and J.E. Gubernatis}
}

@Article{Boninsegni_LowTempPhys_1996,
author={Boninsegni, Massimo
and Ceperley, David M.},
title={Density fluctuations in liquid4He. Path integrals and maximum entropy},
journal={Journal of Low Temperature Physics},
year={1996},
month={Sep},
day={01},
volume={104},
number={5},
pages={339-357},
issn={1573-7357},
doi={10.1007/BF00751861},
url={https://doi.org/10.1007/BF00751861}
}

@article{Asakawa_PPNP_2001,
title = {Maximum entropy analysis of the spectral functions in lattice QCD},
journal = {Progress in Particle and Nuclear Physics},
volume = {46},
number = {2},
pages = {459-508},
year = {2001},
issn = {0146-6410},
doi = {https://doi.org/10.1016/S0146-6410(01)00150-8},
url = {https://www.sciencedirect.com/science/article/pii/S0146641001001508},
author = {M. Asakawa and Y. Nakahara and T. Hatsuda}
}

@article{chuna_PRE_2025,
  title={Estimates of the dynamic structure factor for the finite temperature electron liquid via analytic continuation of path integral Monte Carlo data},
  author={Chuna, Thomas and Barnfield, Nicholas and Vorberger, Jan and Friedlander, Michael P and Hoheisel, Tim and Dornheim, Tobias},
  journal={Physical Review B},
  volume={112},
  number={12},
  pages={125112},
  year={2025},
  publisher={APS}
}

@article{gunnarsson_PRB-MEM_2010,
  title={Analytical continuation of imaginary axis data using maximum entropy},
  author={Gunnarsson, O and Haverkort, MW and Sangiovanni, G},
  journal={Physical Review B—Condensed Matter and Materials Physics},
  volume={81},
  number={15},
  pages={155107},
  year={2010},
  publisher={APS}
}

@article{Kora_PRB_2018,
  title = {Dynamic structure factor of superfluid $^{4}\mathrm{He}$ from quantum Monte Carlo: Maximum entropy revisited},
  author = {Kora, Youssef and Boninsegni, Massimo},
  journal = {Phys. Rev. B},
  volume = {98},
  issue = {13},
  pages = {134509},
  numpages = {8},
  year = {2018},
  month = {Oct},
  publisher = {American Physical Society},
  doi = {10.1103/PhysRevB.98.134509},
  url = {https://link.aps.org/doi/10.1103/PhysRevB.98.134509}
}

@article{rothkopf_arXiv_2012,
  title={Improved maximum entropy method with an extended search space},
  author={Rothkopf, Alexander},
  journal={arXiv preprint arXiv:1208.5162},
  year={2012}
}

@article{asakawa_arXiv_2020,
  title={Comment on ``Heavy Quarkonium in Extreme Conditions'''},
  author={Asakawa, Masayuki},
  journal={arXiv preprint arXiv:2001.10205},
  year={2020}
}

@article{rothkopf_Data_2020,
  title={Bryan’s Maximum Entropy Method—Diagnosis of a Flawed Argument and Its Remedy},
  author={Rothkopf, Alexander},
  journal={Data},
  volume={5},
  number={3},
  pages={85},
  year={2020},
  publisher={MDPI}
}

@article{chuna_JPA_2025,
  title={Dual formulation of the maximum entropy method applied to analytic continuation of quantum Monte Carlo data},
  author={Chuna, Thomas and Barnfield, Nicholas and Dornheim, Tobias and Friedlander, Michael P and Hoheisel, Tim},
  journal={Journal of Physics A: Mathematical and Theoretical},
  volume={58},
  number={33},
  pages={335203},
  year={2025},
  doi={10.1088/1751-8121/adf924}
}

@article{bergeron_PRE_2016,
  title={Algorithms for optimized maximum entropy and diagnostic tools for analytic continuation},
  author={Bergeron, Dominic and Tremblay, A-MS},
  journal={Physical Review E},
  volume={94},
  number={2},
  pages={023303},
  year={2016},
  publisher={APS}
}

@article{rothkopf_PhysRep_2020,
  title={Heavy quarkonium in extreme conditions},
  author={Rothkopf, Alexander},
  journal={Physics Reports},
  volume={858},
  pages={1--117},
  year={2020},
  publisher={Elsevier}
}

@article{fei_PRL_2021,
  title={Nevanlinna analytical continuation},
  author={Fei, Jiani and Yeh, Chia-Nan and Gull, Emanuel},
  journal={Physical Review Letters},
  volume={126},
  number={5},
  pages={056402},
  year={2021},
  publisher={APS}
}

@article{fuchs_PRE_2010,
  title={Analytic continuation of quantum Monte Carlo data by stochastic analytical inference},
  author={Fuchs, Sebastian and Pruschke, Thomas and Jarrell, Mark},
  journal={Physical Review E—Statistical, Nonlinear, and Soft Matter Physics},
  volume={81},
  number={5},
  pages={056701},
  year={2010},
  publisher={APS}
}

@article{kim_JHEP_2018,
  title={Quarkonium in-medium properties from realistic lattice NRQCD},
  author={Kim, Seyong and Petreczky, Peter and Rothkopf, Alexander},
  journal={Journal of High Energy Physics},
  volume={2018},
  number={11},
  pages={1--83},
  year={2018},
  publisher={Springer}
}

@article{kelly_PRD_2018,
  title={Bayesian study of relativistic open and hidden charm in anisotropic lattice QCD},
  author={Kelly, Aoife and Rothkopf, Alexander and Skullerud, Jon-Ivar},
  journal={Physical Review D},
  volume={97},
  number={11},
  pages={114509},
  year={2018},
  publisher={APS}
}

@article{Goulko_PRB_2017,
  title = {Numerical analytic continuation: Answers to well-posed questions},
  author = {Goulko, Olga and Mishchenko, Andrey S. and Pollet, Lode and Prokof'ev, Nikolay and Svistunov, Boris},
  journal = {Phys. Rev. B},
  volume = {95},
  issue = {1},
  pages = {014102},
  numpages = {11},
  year = {2017},
  month = {Jan},
  publisher = {American Physical Society},
  doi = {10.1103/PhysRevB.95.014102},
  url = {https://link.aps.org/doi/10.1103/PhysRevB.95.014102}
}

@article{huang2023acflow,
  title={ACFlow: An open source toolkit for analytic continuation of quantum Monte Carlo data},
  author={Huang, Li},
  journal={Computer Physics Communications},
  volume={292},
  pages={108863},
  year={2023},
  publisher={Elsevier}
}

@article{Kaufmann_CPC-anacont_2023,
title = {ana\_cont: Python package for analytic continuation},
journal = {Computer Physics Communications},
volume = {282},
pages = {108519},
year = {2023},
issn = {0010-4655},
doi = {https://doi.org/10.1016/j.cpc.2022.108519},
url = {https://www.sciencedirect.com/science/article/pii/S0010465522002387},
author = {Josef Kaufmann and Karsten Held},
}

@book{brockwell2002TSA,
  title={Introduction to time series and forecasting},
  author={Brockwell, Peter J and Davis, Richard A},
  year={2002},
  publisher={Springer}
}

@article{wu2007trend,
  title={On the trend, detrending, and variability of nonlinear and nonstationary time series},
  author={Wu, Zhaohua and Huang, Norden E and Long, Steven R and Peng, Chung-Kang},
  journal={Proceedings of the National Academy of Sciences},
  volume={104},
  number={38},
  pages={14889--14894},
  year={2007},
  publisher={National Acad Sciences}
}

@article{adhikari2013TSA,
  title={An introductory study on time series modeling and forecasting},
  author={Adhikari, Ratnadip and Agrawal, Ramesh K},
  journal={arXiv preprint arXiv:1302.6613},
  year={2013}
}

@book{james2013statisticallearning,
  title={An introduction to statistical learning},
  author={James, Gareth and Witten, Daniela and Hastie, Trevor and Tibshirani, Robert and others},
  volume={112},
  year={2013},
  publisher={Springer}
}

@misc{meyer2020biasvariance,
  author       = {Johannes Jakob Meyer},
  title        = {The Multivariate Bias-Variance Decomposition},
  howpublished = {\url{https://johannesjakobmeyer.com/blog/005-multivariate-bias-variance-decomposition/}},
  year         = {2020},
  note         = {Accessed: 2025-10-29}
}

@misc{yeh2019biasvariance,
  author       = {Chris Yeh},
  title        = {Bias-Variance Decomposition of Mean Squared Error},
  howpublished = {\url{https://chrisyeh96.github.io/2019/07/18/bias-variance-decomposition.html}},
  year         = {2019},
  note         = {Last updated December 5, 2022. Accessed: 2025-10-29}
}

@book{berg_book_2004,
  title={Markov chain Monte Carlo simulations and their statistical analysis: with web-based Fortran code},
  author={Berg, Bernd A},
  year={2004},
  publisher={World Scientific Publishing Company}
}

@article{dornheim_PRL-DLFC_2018,
  title={Ab initio path integral Monte Carlo results for the dynamic structure factor of correlated electrons: From the electron liquid to warm dense matter},
  author={Dornheim, Tobias and Groth, Simon and Vorberger, Jan and Bonitz, Michael},
  journal={Physical Review Letters},
  volume={121},
  number={25},
  pages={255001},
  year={2018},
  publisher={APS}
}

@article{chuna_JCP_2025,
  title={Second roton feature in the strongly coupled electron liquid},
  author={Chuna, Thomas M and Vorberger, Jan and Tolias, Panagiotis and Benedix Robles, Alexander and Hecht, Michael and Hofmann, Phil-Alexander and Moldabekov, Zhandos A and Dornheim, Tobias},
  journal={The Journal of Chemical Physics},
  volume={163},
  number={3},
  year={2025},
  publisher={AIP Publishing}
}

@article{Dornheim_MRE_2023,
  author = {Dornheim, Tobias and Moldabekov, Zhandos and Tolias, Panagiotis and Böhme, Maximilian and Vorberger, Jan},  
  title = {Physical insights from imaginary-time density--density correlation functions},
  journal = {Matter and Radiation at Extremes},
  volume = {8},
  pages = {056601},
  year = {2023},
  publisher = {American Physical Society},
  doi = {10.1063/5.0149638},
  url = {https://doi.org/10.1063/5.0149638}
}

@article{dornheim_JCP-MLstatic_2019,
author = {T. Dornheim and J. Vorberger and S. Groth and N. Hoffmann and Zh.A. Moldabekov and M. Bonitz},
journal = {J. Chem. Phys},
pages = {194104},
title = {The Static Local Field Correction of the Warm Dense Electron Gas: An ab Initio Path Integral {M}onte {C}arlo Study and Machine Learning Representation},
volume = {151},
year = {2019},
url = {https://aip.scitation.org/doi/full/10.1063/1.5123013},
}

@article{dornheim_PRB-analyticstatic_2021,
  title={Analytical representation of the local field correction of the uniform electron gas within the effective static approximation},
  author={Dornheim, Tobias and Moldabekov, Zhandos A and Tolias, Panagiotis},
  journal={Physical Review B},
  volume={103},
  number={16},
  pages={165102},
  year={2021},
  publisher={APS}
}

@article{Filinov_PRA_2012,
  title = {Collective and single-particle excitations in two-dimensional dipolar Bose gases},
  author = {Filinov, A. and Bonitz, M.},
  journal = {Phys. Rev. A},
  volume = {86},
  issue = {4},
  pages = {043628},
  numpages = {26},
  year = {2012},
  month = {Oct},
  publisher = {American Physical Society},
  doi = {10.1103/PhysRevA.86.043628},
  url = {https://link.aps.org/doi/10.1103/PhysRevA.86.043628}
}

@book{anderson_QMCtextbook_2007,
  title={Quantum Monte Carlo: Origins, Development, Applications},
  author={Anderson, J.B.},
  isbn={9780195310108},
  lccn={2006040145},
  url={https://books.google.de/books?id=\_QUSDAAAQBAJ},
  year={2007},
  publisher={Oxford University Press, USA}
}

@article{Dornheim_JCP_ITCF_2021,
author = {Dornheim,Tobias  and Moldabekov,Zhandos A.  and Vorberger,Jan },
title = {Nonlinear density response from imaginary-time correlation functions: Ab initio path integral Monte Carlo simulations of the warm dense electron gas},
journal = {The Journal of Chemical Physics},
volume = {155},
number = {5},
pages = {054110},
year = {2021},
doi = {10.1063/5.0058988},
URL = {https://doi.org/10.1063/5.0058988}
}

@Article{Dornheim_SciRep_2022,
author={Dornheim, Tobias
and Moldabekov, Zhandos A.
and Vorberger, Jan
and Militzer, Burkhard},
title={Path integral Monte Carlo approach to the structural properties and collective excitations of liquid $^3$He without fixed nodes},
journal={Scientific Reports},
year={2022},
month={Jan},
day={13},
volume={12},
number={1},
pages={708},
issn={2045-2322},
doi={10.1038/s41598-021-04355-9},
url={https://doi.org/10.1038/s41598-021-04355-9}
}

@article{Foulkes_RMP_2001,
  title = {Quantum Monte Carlo simulations of solids},
  author = {Foulkes, W. M. C. and Mitas, L. and Needs, R. J. and Rajagopal, G.},
  journal = {Rev. Mod. Phys.},
  volume = {73},
  issue = {1},
  pages = {33--83},
  numpages = {0},
  year = {2001},
  month = {Jan},
  publisher = {American Physical Society},
  doi = {10.1103/RevModPhys.73.33},
  url = {https://link.aps.org/doi/10.1103/RevModPhys.73.33}
}

@article{Ceperley_RevModPhys_1995,
author = {D. M. Ceperley},
journal = {Rev. Mod. Phys},
pages = {279},
title = {Path integrals in the theory of condensed helium},
volume = {67},
year = {1995},
url = {https://journals.aps.org/rmp/abstract/10.1103/RevModPhys.67.279},
}

@article{siegfried_review,
author = {S. H. Glenzer and R. Redmer},
journal = {Rev. Mod. Phys},
pages = {1625},
title = {X-ray Thomson scattering in high energy density plasmas},
volume = {81},
year = {2009},
url = {https://journals.aps.org/rmp/abstract/10.1103/RevModPhys.81.1625},
}

@Article{Boninsegni1996,
author={Boninsegni, Massimo
and Ceperley, David M.},
title={{Density fluctuations in liquid $^4$He. Path integrals and maximum entropy}},
journal={Journal of Low Temperature Physics},
year={1996},
month={Sep},
day={01},
volume={104},
number={5},
pages={339-357},
issn={1573-7357},
doi={10.1007/BF00751861},
url={https://doi.org/10.1007/BF00751861}
}

@article{Schoerner_PRE_2023,
  title = {X-ray Thomson scattering spectra from density functional theory molecular dynamics simulations based on a modified Chihara formula},
  author = {Sch\"orner, Maximilian and Bethkenhagen, Mandy and D\"oppner, Tilo and Kraus, Dominik and Fletcher, Luke B. and Glenzer, Siegfried H. and Redmer, Ronald},
  journal = {Phys. Rev. E},
  volume = {107},
  issue = {6},
  pages = {065207},
  numpages = {14},
  year = {2023},
  month = {Jun},
  publisher = {American Physical Society},
  doi = {10.1103/PhysRevE.107.065207},
  url = {https://link.aps.org/doi/10.1103/PhysRevE.107.065207}
}

@article{Rabani_PNAS_2002,
author = {Eran Rabani  and David R. Reichman  and Goran Krilov  and Bruce J. Berne },
title = {The calculation of transport properties in quantum liquids using the maximum entropy numerical analytic continuation method: Application to liquid para-hydrogen},
journal = {Proceedings of the National Academy of Sciences},
volume = {99},
number = {3},
pages = {1129-1133},
year = {2002},
doi = {10.1073/pnas.261540698},
URL = {https://www.pnas.org/doi/abs/10.1073/pnas.261540698},
eprint = {https://www.pnas.org/doi/pdf/10.1073/pnas.261540698}
}

@article{Dornheim_review,
 author = {Dornheim,Tobias  and Moldabekov,Zhandos A.  and Ramakrishna,Kushal  and Tolias,Panagiotis  and Baczewski,Andrew D.  and Kraus,Dominik  and Preston,Thomas R.  and Chapman,David A.  and Böhme,Maximilian P.  and Döppner,Tilo  and Graziani,Frank  and Bonitz,Michael  and Cangi,Attila  and Vorberger,Jan },
title = {Electronic density response of warm dense matter},
journal = {Physics of Plasmas},
volume = {30},
number = {3},
pages = {032705},
year = {2023},
doi = {10.1063/5.0138955},

URL = { 
        https://doi.org/10.1063/5.0138955
    
}

}

@article{boninsegni1,
author = {M. Boninsegni and N. V. Prokofev and B. V. Svistunov},
journal = {Phys. Rev. E},
pages = {036701},
title = {Worm algorithm and diagrammatic {M}onte {C}arlo: A new approach to continuous-space path integral {M}onte {C}arlo simulations},
volume = {74},
year = {2006},
url = {https://journals.aps.org/pre/abstract/10.1103/PhysRevE.74.036701},
}

@article{Ferre_PRB_2016,
  title = {Dynamic structure factor of liquid $^{4}\mathrm{He}$ across the normal-superfluid transition},
  author = {Ferr\'e, G. and Boronat, J.},
  journal = {Phys. Rev. B},
  volume = {93},
  issue = {10},
  pages = {104510},
  numpages = {9},
  year = {2016},
  month = {Mar},
  publisher = {American Physical Society},
  doi = {10.1103/PhysRevB.93.104510},
  url = {https://link.aps.org/doi/10.1103/PhysRevB.93.104510}
}

\end{document}